\documentclass{article}

\usepackage[nonatbib,preprint]{neurips_2020}

\usepackage[utf8]{inputenc} 
\usepackage[T1]{fontenc}    
\usepackage{hyperref}       
\usepackage{url}            
\usepackage{booktabs}       
\usepackage{amsfonts}       
\usepackage{nicefrac}       
\usepackage{microtype}      
\usepackage{cite}

\usepackage{graphicx,subcaption}        
\usepackage{algorithm,algpseudocode}    
\usepackage{amsmath,amssymb}            
\usepackage{multirow}
\usepackage[table,xcdraw]{xcolor}
\usepackage[font=footnotesize,labelfont=bf]{caption}

\usepackage{xcolor}         

\title{Learning Illumination Patterns for Coded Diffraction Phase Retrieval}
\author{%
  Zikui Cai  \\
  ECE Department \\
  UC Riverside\\
  \texttt{zcai032@ucr.edu}
  \And
  Rakib Hyder \\
  ECE Department \\
  UC Riverside\\
  \texttt{rhyde001@ucr.edu}
  \And
   M. Salman Asif \\ 
   ECE Department\\
  UC Riverside\\
  \texttt{sasif@ece.ucr.edu} 
}

\begin{document}
\maketitle

\begin{abstract}

Signal recovery from nonlinear measurements involves solving an iterative optimization problem. In this paper, we present a framework to optimize the sensing parameters to improve the quality of the signal recovered by the given iterative method. In particular, we learn illumination patterns to recover signals from coded diffraction patterns using a fixed-cost alternating minimization-based phase retrieval method. 
Coded diffraction phase retrieval is a physically realistic system in which the signal is first modulated by a sequence of codes before the sensor records its Fourier amplitude. We represent the phase retrieval method as an unrolled network with a fixed number of layers and minimize the recovery error by optimizing over the measurement parameters. Since the number of iterations/layers are fixed, the recovery incurs a fixed cost. 
We present extensive simulation results on a variety of datasets under different conditions and a comparison with existing methods. Our results demonstrate that the proposed method provides near-perfect reconstruction using patterns learned with a small number of training images. Our proposed method provides significant improvements over existing methods both in terms of accuracy and speed.
\end{abstract}



\section{Introduction}
The problem of signal recovery from nonlinear measurements arises in various imaging and signal processing tasks \cite{shechtman2015phase,maiden2009improved,millane1990phase,rodenburg2008ptychography,candes2015phasediff}. Conventional methods for solving such inverse problems use an iterative method to recover the signal from given measurements. In this paper, we present a framework to optimize over the measurement parameters to improve the quality of signals recovered by the given iterative method. In particular, we learn illumination patterns to recover the signal from coded diffraction patterns (CDP) using a fixed-cost alternating minimization method. 

Coded diffraction imaging is a specific instance of Fourier phase retrieval problems. Phase retrieval refers to a broad class of nonlinear inverse problems where we seek to recover a complex- (or real-) valued signal from its phase-less (or sign-less) measurements \cite{fienup1978reconstruction,candes2013phaselift,fienup1982phase,jaganathan2015phase,shechtman2015phase,muminov2020small}. In practice, these problems often arise in coherent optical imaging where an image sensor records the intensity of the Fourier measurements of the object of interest. In coded diffraction imaging, the signal of interest gets modulated by a sequence of known illumination patterns/masks before observing the Fourier intensity at the sensor \cite{candes2013phaselift,jaganathan2015phase,shechtman2015phase}.
Applications include X-ray crystallography \cite{watson1953structure,millane1990phase,harrison1993phase}, astronomy \cite{fienup1987phase,gonsalves2014perspectives}, microscopy \cite{misell1973method,miao2008extending,rodenburg2008ptychography,tian2014multiplexed}, speech processing and acoustics \cite{rabiner1993fundamentals,balan2006signal,balan2010signal,jaganathan2016stft}, and quantum mechanics \cite{corbett2006pauli,reichenbach1998philosophic}.

We can model the sensor measurements for coded diffraction imaging as follows. Let us denote the signal of interest as $x \in \mathbb{R}^n$ or $\mathbb{C}^n$ that is modulated by $T$ illumination patterns $D = \{d_1, \ldots, d_T\}$, where $d_t \in \mathbb{R}^n$ or $\mathbb{C}^n$. The amplitude of sensor measurements for $t$th illumination pattern can be written as 
\begin{equation}
    y_t =  |\mathcal{F} (d_t \odot x)|,
    \label{eq:yt}
\end{equation}
where $\mathcal{F}$ denotes the Fourier transform operator and $\odot$ denotes an element-wise product. 
We note that real sensor measurements are proportional to the intensity of the incoming signal (i.e., square of the Fourier transform). In practice, however, solving the inverse problem with (non-square) amplitude measurements provides better results; therefore, we use the amplitude measurements throughout the paper.

\begin{figure}[!th]
    \centering
    \includegraphics[width=1\columnwidth]{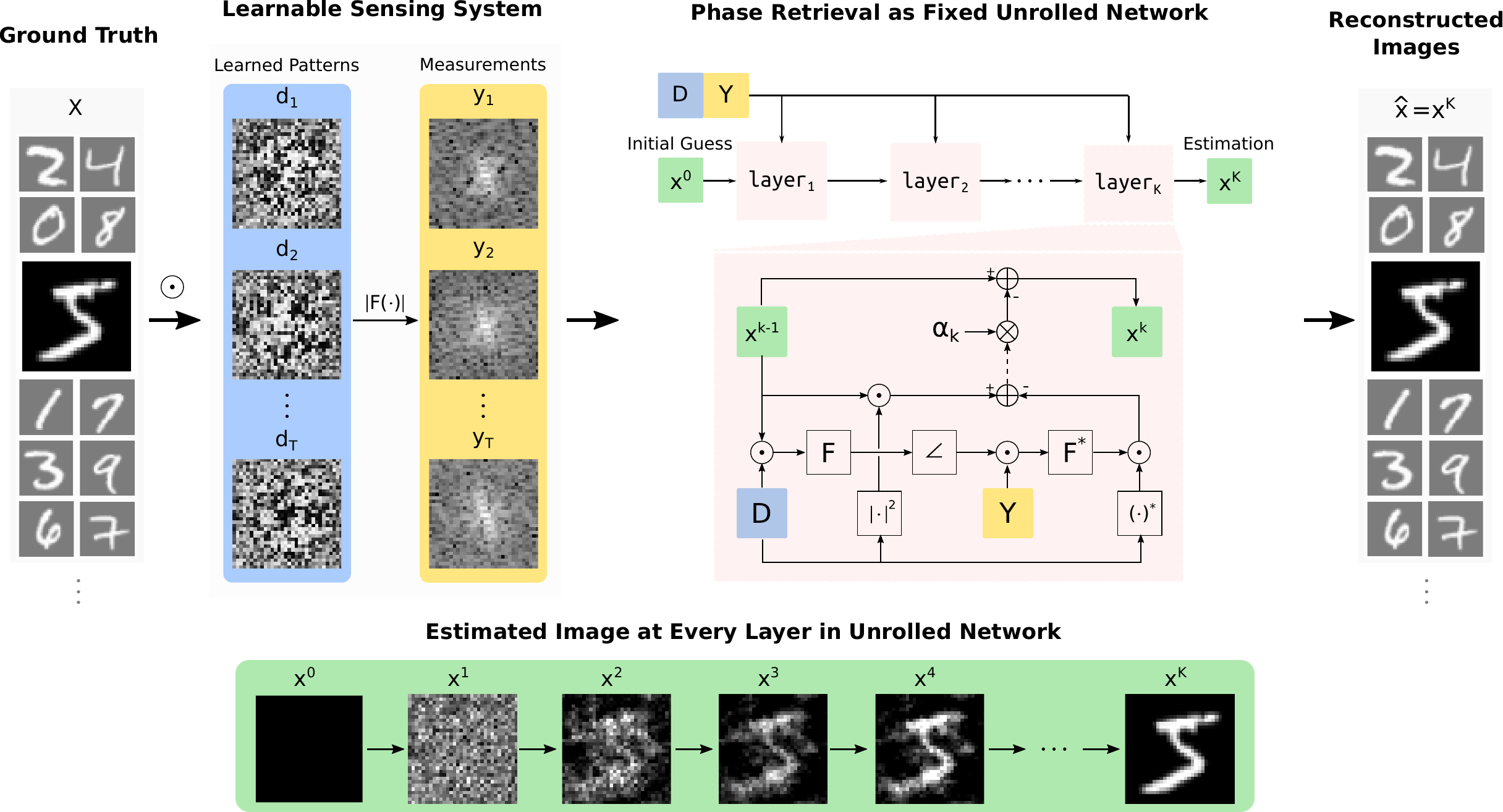}
    \caption{The block diagram of our proposed framework to learn illumination patterns while solving coded diffraction imaging using $K$ iterations of a phase retrieval algorithm. The iterative algorithm is represented as an unrolled network with $K$ layers. Steps at every iteration are fixed and depicted as an unrolled network (details can be found in Algorithm~\ref{algo:test}). Phase retrieval algorithm uses the measurements $Y = \{y_t\}$ and  illumination patterns $D = \{d_t\}$ to provide an estimate $x^K$ after $K$ iterations. Our goal is to learn the illumination patterns $D$ to minimize the error between the estimated $x^K$ and the ground truth.}
    \label{fig:intro}
\end{figure}

To recover the signal $x$ from the nonlinear measurements, we can solve the following optimization problem: 
\begin{equation}\label{eq:PR}
    \min_x \; \sum_{t=1}^T \|y_t - |\mathcal{F}(d_t \odot x)| \|_2^2.
\end{equation}
In recent years, a number of iterative algorithms have been proposed for solving the problem in \eqref{eq:PR}, which includes lifting-based convex methods, alternating minimization-based nonconvex methods, and greedy methods \cite{candes2013phaselift,gross2017improved,netrapalli2013phase,hyder2019alternating,Jagatap2017}.

Our goal is to learn a set of illumination patterns to optimize the recovery of an alternating minimization (AltMin) algorithm for solving the problem in \eqref{eq:PR}. The AltMin method can be viewed as an unrolled gradient descent network, as shown in Fig.~\ref{fig:intro}, where we fix the steps at every iteration and the total number of iterations for AltMin. One forward pass through the unrolled network is equivalent to $K$ iterations of the AltMin algorithm. To minimize the computational complexity of the recovery algorithm, we keep the total number of iterations very small (e.g., $K=50$). At the training stage, we optimize over the illumination patterns to minimize the error between the AltMin outputs after $K$ iterations and the ground truth training images. At the test time, we solve the problem in \eqref{eq:PR} using $K$ AltMin iteration with the learned illumination patterns (equivalent to one forward pass). 
We evaluated our method on different image datasets and compared against existing methods for coded diffraction imaging. We demonstrate that our proposed method of designing illumination patterns for a fixed-cost algorithm outperforms existing methods both in terms of accuracy and speed.

\section{Related Work}

\textbf{Phase Retrieval and Coded Diffraction Patterns:} Fourier phase retrieval problem arises in a number of imaging systems because standard image sensors can only record intensity of the observed measurements. This problem has been extensively studied over last five decades in optics, signal processing, and optimization \cite{gerchberg1972practical, fienup1982phase, millane1990phase, rodenburg2008ptychography,maiden2009improved, shechtman2015phase, hand2018phase,nayer2019phaseless}. Coded diffraction imaging is a physically realistic setup in which we can first modulate by signal of interest and then collect the intensity measurements \cite{candes2015phasediff,chandra2017phasepack}. The modulation can be performed using a spatial light modulator or custom transparencies \cite{miao2008extending,rodenburg2008ptychography,ptychTCI}. The recovery problems involves solving a phase retrieval problem; the presence of modulation patterns makes this a more tractable problem compared to classical Fourier phase retrieval \cite{candes2015phasediff}. 

The algorithms for solving phase retrieval problem can be broadly divided into non-convex and convex methods. Classical algorithms for phase retrieval rely on solving the underlying non-convex problem using alternating minimization \cite{fienup1982phase,bauschke2002phase,netrapalli2013phase}. Amplitude flow \cite{wang2016sparse,wang2016solving},  Wirtinger flow \cite{candes2015phase,zhang2016reshaped,chen2015solving,cai2016optimal}, alternating minimization (AltMin) \cite{netrapalli2013phase,zhang2016reshaped,Jagatap2017,hyder2019alternating} are recent methods that solve the non-convex problem.
Convex methods usually lift the nonconvex problem of signal recovery from quadratic measurements into a convex problem of low-rank matrix recovery from linear measurements  
\cite{candes2013phaselift,soltani2016fast,fazel2008compressed}.
The PhaseLift algorithm \cite{candes2013phaselift} and its variations \cite{gross2017improved, candes2015phasediff} can be considered under this class. 
We can also incorporate prior knowledge about the signal structure (e.g., sparsity, support, or positivity) in the recovery process  constraints \cite{ohlsson2012cprl,li2013sparse,bahmani2015efficient,jaganathan2012recovery,netrapalli2013phase,cai2016optimal,wang2016sparse}. 

\textbf{Data-Driven Approaches for Phase Retrieval:} 
A number of papers have recently explored the idea of replacing the classical (hand-designed) signal priors with deep generative priors for solving inverse problems \cite{bora2017compressed,hand2016compressed,ulyanov2018deep,van2018compressed}. Some of the generative prior-based approaches for phase retrieval are presented in \cite{hyder2019alternating,jagatap2019algorithmic, hand2018phase,shamshad2018robust,metzler2020deep}. 

Another growing trend is to apply deep learning to solve inverse problems (including phase retrieval) in an end-to-end manner, where deep networks are trained to learn a mapping from sensor measurements to the signal of interest using a large number of measurement-signal pairs. A few examples demonstrating the benefit of the data-driven approaches include robust phase retrieval \cite{metzler2018prdeep}, Fourier ptychographic microscopy \cite{kellman2019data}, holographic image reconstruction  \cite{rivenson2018phase}, and correlography for non-line-of-sight imaging \cite{metzler2020deep2}. 

While our method is partially driven by data, our goal is not to learn a signal prior or a mapping from measurements to signal. We use data to learn the illumination patterns for a fixed recovery algorithm. The number of training images required by our method is extremely small (32 or 128 images only). Furthermore, the patterns we learn on one class of images provide good results on other types of images (see Table~\ref{table:generalzation}). Apart from the great flexibility, our method uses a well-defined AltMin routine, where we know exact steps for every iteration as opposed to the black-box deep models.

\textbf{Unrolled Network for Inverse Problem:}
Iterative methods for solving the inverse problems, such as AltMin or other first-order methods, can be represented as unrolled networks. Every layer of such a network performs the same steps as a single iteration of the original method  \cite{kellman2019physics,diamond2017unrolled,gregor2010learning,wang2016proximal,hammernik2018learning,sun2016deep,kamilov2016learning,bostan2018learning,monga2019algorithm,liang2019deep}. Some parameters of the iterative steps can be learned from data (e.g., step size, denoiser, or threshold parameters) but the basic structure and physical forward model are kept intact.

\textbf{Learn to Sense:} Deep learning methods have also been recently used to design the sensing system; especially in the context of compressive sensing and computational imaging \cite{mousavi2017learning,wu2019learning, Bergman:2020:DeepLiDAR,wang2020learning}. The main objective in these methods is similar to ours, which is to select sensor parameters to recover best possible signal/image from the sensor measurements. The sensor parameters may involve selection of samples/frames, design of sampling waveforms, or illumination patters as we discuss in this paper. In principle, the sensor can be treated as the first layer of the network with some physical constraints on the parameters \cite{kellman2019physics}. In contrast to most of the existing methods that learn a deep network to solve the inverse problem, our method uses a predefined iterative method as an unrolled network while learning the illumination patterns using a small number of training images. 

\section{Proposed Method}
We use $N$ training images ($x_1,\ldots, x_N$) to learn $T$ illumination patterns that provide best reconstruction using a predefined (iterative) phase retrieval algorithm. Furthermore, to ensure that the illumination patterns are physically realizable, we constrain their values to be in the range $[0,1]$. We use a sigmoid function over unconstrained parameters $\Theta = \{\theta_1, \ldots, \theta_T\}$ to define the illumination patterns; that is, $d_t = \text{sigmoid}(\theta_t)$ for all $t=1,\ldots,T$. 

Our proposed method for learning illumination patterns can be divided into two parts: The first (inner) part involves solving the phase retrieval problem with given coded diffraction patterns using AltMin as an unrolled network (see block diagram in Fig.~\ref{fig:intro}); Second part is updating the illumination patterns based on backpropagating the image reconstruction loss. These two parts together yield optimal image reconstruction and illumination patterns. Pseudocodes for both parts are listed in Algorithms~\ref{algo:train},\ref{algo:test}.

\begin{algorithm}[tb]
   \caption{Learning illumination patterns}
   \label{algo:train}
\begin{algorithmic}
   \State {\bfseries Input:} Training set $X$ with $N$ images $X=\{x_1, \ldots, x_N\}$.
   
   \State {\bfseries Initialize: } Initialize the optimization variables for $T$ patterns as $\Theta = \{\theta_1,\ldots, \theta_T \}$ from a uniform distribution. 
   \For{$\text{epoch} = 1, 2, ..., M$}         \Comment{M epochs}
        \State Generate illumination patterns in the range [0,1] as $d_t = \text{sigmoid}(\theta_t)$ for all $t$.
        \For{$n=1, 2, ..., N$}      \Comment{N samples}
            \State $Y^n = \{y_1^n, \ldots, y_T^n ~|~y_t^n=|\mathcal{F} (d_t \odot x_n)| \}$
            \State $x_n^K(\Theta) \gets \texttt{solveCDP($Y^n$,$D$)}$ \Comment{\texttt{solveCDP} for each sample}
        \EndFor
        \State $L_\Theta = \sum_{n=1}^{N} \| x_n - x_n^K(\Theta)\|_2^2$
    	\State $\Theta \leftarrow \Theta - \beta \nabla_{\Theta}L_{\Theta}$ \Comment{Optimize over $\Theta$}
     \EndFor
     \State {\bfseries Output:} Optimal illumination patterns ${D} = \{d_1, \ldots, d_T ~|~ d_t = \text{sigmoid}(\theta_t)\}$.
\end{algorithmic}
\end{algorithm}

\begin{algorithm}[tb]
   \caption{\texttt{solveCDP($Y,D$)} via AltMin}
   \label{algo:test}
\begin{algorithmic}
   \State {\bfseries Input:}  Measurements $Y = \{y_1, \ldots, y_t\}$ and illumination patterns $D =\{d_1,\ldots, d_T\}$.
   
   \State {\bfseries Initialization:}  Initialize estimate $x^0 = 0$.
	\For{$k = 1, 2, ..., K$}    \Comment{K iterations of AltMin}
        \State $p_t^{k-1} \gets \text{sign}(\mathcal{F}(d_t \odot x^{k-1})$ 
        %
        %
        \State $\nabla_x L_{x,p} = \frac{2}{T}\sum_{t=1}^T [|d_t|^2\odot x^{k-1} - d_t^* \odot \mathcal{F}^*(p_t^{k-1}\odot y_t)]$
    	\State $x^k \leftarrow x^{k-1} - \alpha\nabla_{x}L_{x,p} $ 
	\EndFor
   \State {\bfseries Output:} Estimated signal $x^K$ 

\end{algorithmic}
\end{algorithm}

\textbf{Phase retrieval as alternating minimization (AltMin):} Given measurements $Y = \{y_1,\ldots, y_T\}$ and illumination patterns $D = \{d_1,\ldots,d_T\}$, we seek to solve the CDP phase retrieval problem by minimizing the loss function defined in \eqref{eq:PR} as 
\begin{equation}\label{eq:Lx}
    L_{x} = \frac{1}{T}\sum_{t=1}^{T} \|  y_t - |\mathcal{F} (d_t \odot x)| \|_2^2.
\end{equation}
Even though the loss function in \eqref{eq:Lx} is nonconvex and nonsmooth with respect to $x$, we can minimize it using the well-known alternating minimization (AltMin) with a single gradient descent step \cite{netrapalli2013phase,zhang2016reshaped}. We define a new variable for the  estimated phase of linear measurements as $p_t = \text{phase}(\mathcal{F} (d_t \odot x))$ and reformulate the loss function in \eqref{eq:Lx} into 
\begin{equation}\label{eq:Lx_p}
    L_{x,p} = \frac{1}{T}\sum_{t=1}^{T} \|  p_t \odot y_t - \mathcal{F} (d_t \odot x) \|_2^2.
\end{equation}
The gradient step can be computed as 
\begin{equation}
    \nabla_{x}L_{x,p} = \frac{2}{T}\sum_{t=1}^{T}  |d_t|^2 \odot x - d_t^* \odot \mathcal{F}^*(p_t \odot  y_t). \label{eqn:grad_x}
\end{equation}

We initialize the image estimate as a zero vector and update the estimate at every iteration as $  x^k = x^{k-1} - \alpha_{k-1}\nabla_{x}L_{x,p},$
where $\alpha_{k-1}$ denotes the step size. In our implementation, we used a fixed step size $\alpha$ for all iterations. 
We use $K$ iterations of the gradient descent in the unrolled network and denote the final estimate as $x^K$. We summarize the steps of CDP phase retrieval routine as \texttt{solveCDP} in Algorithm~\ref{algo:test}.

\textbf{Learning illumination patterns:} To learn a set of illumination patterns that provide the best reconstruction with the predefined iterative method (or the unrolled network), we seek to minimize the difference between the original training images and their estimates. In this regard, we minimize the following quadratic loss function with respect to $\Theta$:  
\begin{equation}\label{eq:L_theta}
    L_\Theta = \sum_{n=1}^N \|x_n - x_n^K(\Theta)\|_2^2, 
\end{equation}
where $x_n^K(\Theta)$ denotes the \texttt{solveCDP} estimate of $n$th training image for the given values of $\Theta$. 
Note that for a given values of $\Theta$, we can define illumination patterns as $d_t = \text{sigmoid}(\theta_t)$ and sensor measurements for $x_n$ as $y^n_t = |\mathcal{F}(d_t\odot x_n)|$ for $t=1,\ldots, T$ and $n = 1,\ldots, N$. 
We use Adam optimizer in PyTorch \cite{kingma2014adam,paszke2019pytorch} to minimize the loss function in \eqref{eq:L_theta}. A summary of the algorithm for learning the illumination patterns is also listed in Algorithm \ref{algo:train}.

\section{Experiments}
\textbf{Datasets.}
\label{section:dataset}
We used MNIST digits, Fashion MNIST (F. MNIST), CIFAR10, SVHN and CelebA datasets for training and testing in our experiments. We use 128 images from each of the datasets for training and another 1000 images for testing. To make the tiny-image datasets uniform, we reshaped all of them to $32\times32$ size with grayscale values. Images in CelebA dataset have $218\times178$ pixels, we first converted all the images to grayscale, cropped $178\times178$ region in the center, and resized to $200\times200$. In Fig. \ref{fig:classical}, we evaluated the performance of our method on images used in \cite{metzler2018prdeep}.
 
\textbf{Measurements.} We use the amplitude of the 2D Fourier transform of the images modulated with $T$ illumination patterns as the measurements. Unless otherwise mentioned, we used noiseless measurements. We report results for measurements with Gaussian and Poisson noise in Fig.~\ref{fig:test_noise}.

\textbf{Computing Infrastructure.} Intel Core i7-8700 CPU and NVIDIA TITAN Xp GPU.

\subsection{Setup and hyper-parameter search}
The hyper-parameters include the number of iterations ($K$), step size $\alpha$, and the number of training samples $N$. 
We set the default value of $K=50$, but we will show later that $K$ can be adjusted as a trade-off between better reconstruction quality and shorter run time. 
We tested all methods for $T = \{2,4,8\}$ to evaluate three cases where signal recovery is hard, moderate, and easy, respectively. Through grid search, we found that it provides the best results over all datasets when $\alpha*T = 4$. We also studied the effect of the number of training images and found that illumination patterns learned on $32$ randomly selected images provide good recovery over the entire dataset. The test accuracy improves slightly as we increase the number of training samples. To be safe, we used 128 training images in all our experiments. 

\subsection{Comparison between random and learned illumination patterns}
To demonstrate the advantages of our learned illumination patterns, we compare the performance of learned and random illumination patterns on five different datasets. We learn a set of $T=\{2,4,8\}$ illumination patterns on 128 training images from a dataset and test them on 1000 test images from the same dataset. For random patterns, we draw $T$ independent patterns from Uniform(0,1) distribution and test their performance on the same 1000 samples that we used for the learned case. We repeat this process 30 times and choose the best result to compare with the results for the learned illumination patterns. The average PSNR over all 1000 test image reconstructions is presented in Table~\ref{table:compare_learned_random}, which shows that the learned illumination patterns perform significantly better than the random patterns for all values of $T$. In addition to that, we can observe a transition in the performance for $T=4$, where random patterns provide poor quality reconstructions and learned patterns provide very high quality reconstructions.

\begin{figure*}[!ht]
    \centering 
    \begin{subfigure}[t]{0.24\linewidth}
    \includegraphics[width=1\linewidth]{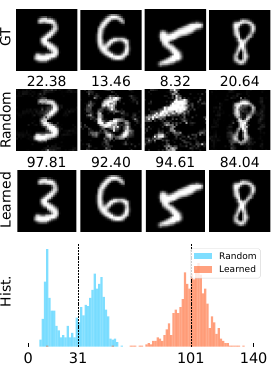}
    \caption{MNIST}
    \end{subfigure}
    \hfill 
    \begin{subfigure}[t]{0.24\linewidth}
    \includegraphics[width=1\linewidth]{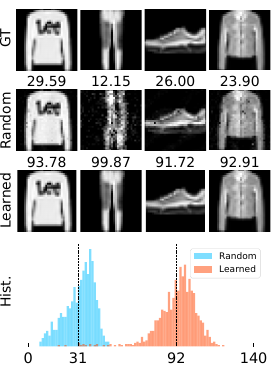}
    \caption{F. MNIST}
    \end{subfigure}
    \hfill
    \begin{subfigure}[t]{0.24\linewidth}
    \includegraphics[width=1\linewidth]{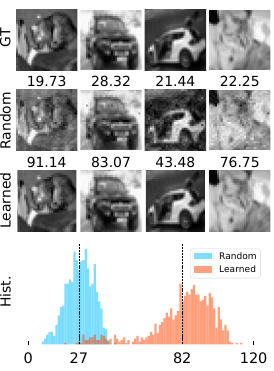}
    \caption{CIFAR10}
    \end{subfigure}
    \hfill
    \begin{subfigure}[t]{0.24\linewidth}
    \includegraphics[width=1\linewidth]{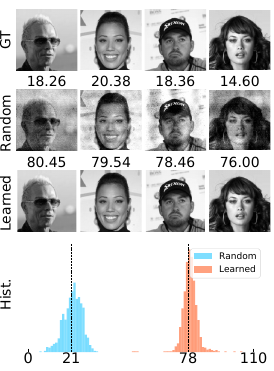}
    \caption{CelebA}
    \end{subfigure}
    \caption{Selected ground truth (GT) images, corresponding reconstructed images using random and learned illumination patterns. PSNR is shown on top of every reconstruction. Below each dataset, we show the histograms of the PSNRs of all images with random patterns (shown in blue) and learned patterns (shown in orange). The dashed vertical line indicates the mean of all PSNRs. We used $T=4$ illumination patterns. Random illumination patterns are selected best out of 30 trials. The learned illumination patterns are trained on 128 training images.}
    \label{fig:compare}
\end{figure*}

To highlight this effect, we show a small set of reconstructed images and histograms of PSNRs of all the reconstructed images from learned and random illumination patterns in Fig.~\ref{fig:compare} for $T=4$ patterns. The result suggests that the learned illumination patterns demonstrate consistently better performance compared to random illumination patterns. 

\begin{table*}[t]
\caption{Average PSNR for learned and random illumination patterns tested on different datasets.}
\label{table:compare_learned_random}
\vskip 0.15in
\begin{center}
\begin{small}
\begin{tabular}{@{}ccccccc@{}}
\toprule
\multirow{2}{*}{PSNR}      & \multicolumn{2}{c}{2 Illumination Patterns} & \multicolumn{2}{c}{4 Illumination Patterns} & \multicolumn{2}{c}{8 Illumination Patterns} \\ \cmidrule(l){2-7} 
         & Random & Learned & Random & Learned & Random & Learned \\ \midrule
MNIST    & 13.59  & 27.77   & 28.61  & 101.50  & 46.43  & 111.79  \\
F. MNIST & 16.26  & 25.17   & 30.05  & 92.44   & 52.05  & 108.85  \\
CIFAR10  & 13.40  & 22.93   & 27.51  & 82.02   & 60.64  & 104.61  \\
SVHN     & 12.24  & 20.27   & 26.24  & 84.43   & 65.33  & 109.25  \\
\multicolumn{1}{l}{CelebA} & 12.67        & 18.49        & 21.67        & 78.98        & 41.49        & 90.29        \\ \bottomrule
\end{tabular}

\end{small}
\end{center}
\vskip -0.1in
\end{table*}

\subsection{Effect of number of iterations/layers (K)}
\begin{figure}[!ht]
    \centering 
    \begin{subfigure}[t]{0.24\linewidth}
    \includegraphics[width=1\linewidth]{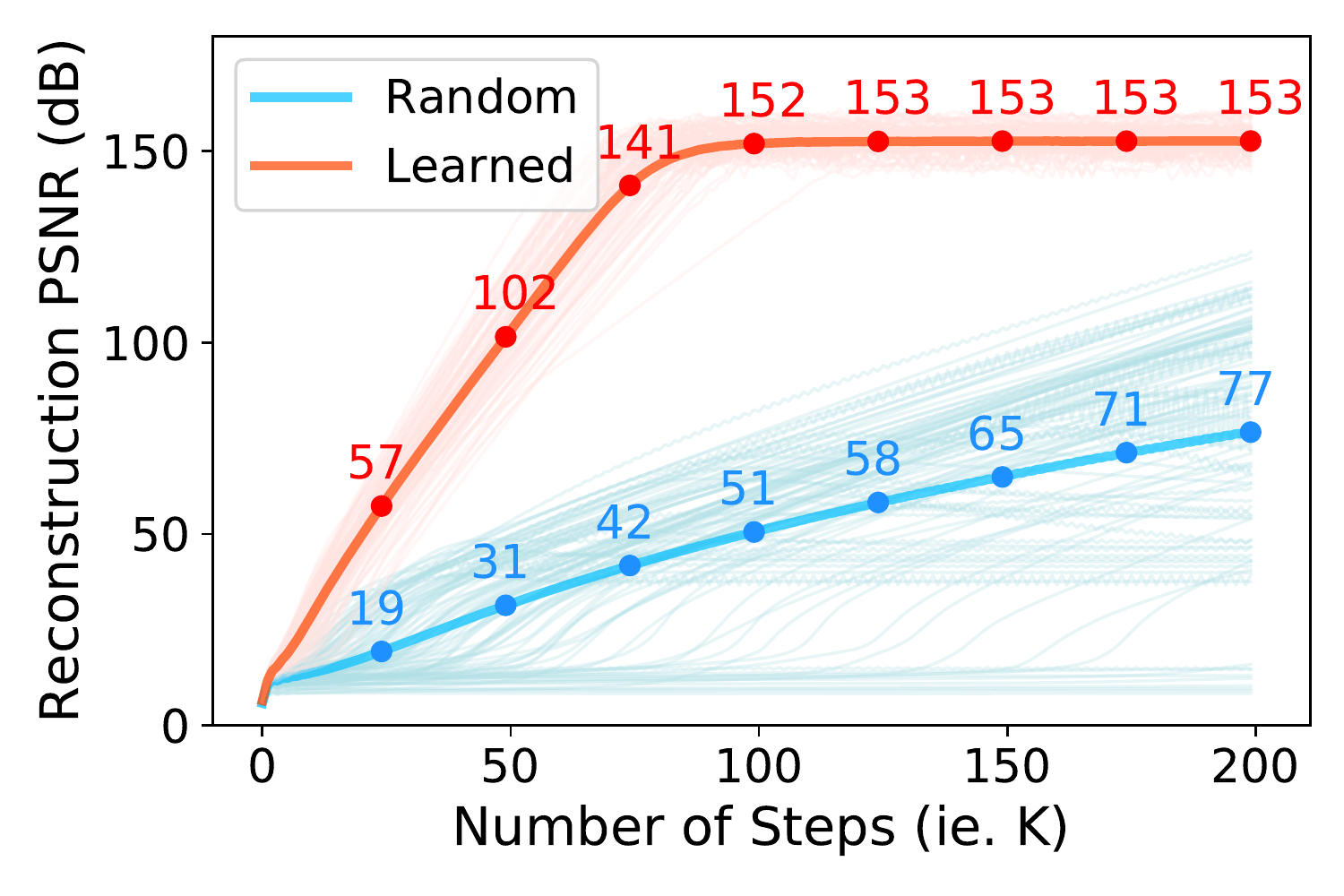}
    \caption{MNIST}
    \end{subfigure}
    \hfill 
    \begin{subfigure}[t]{0.24\linewidth}
    \includegraphics[width=1\linewidth]{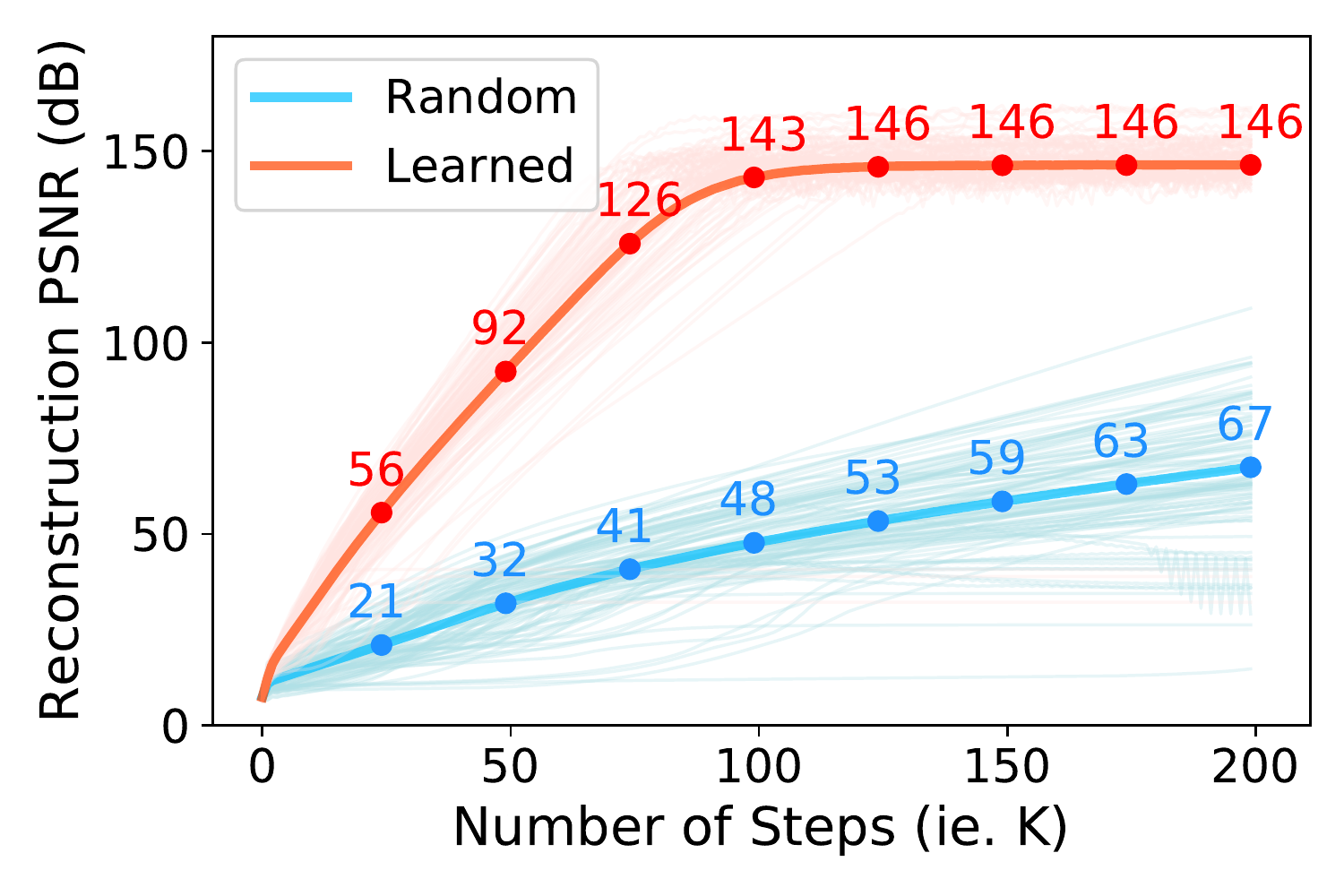}
    \caption{F. MNIST}
    \end{subfigure}
    \hfill
    \begin{subfigure}[t]{0.24\linewidth}
    \includegraphics[width=1\textwidth]{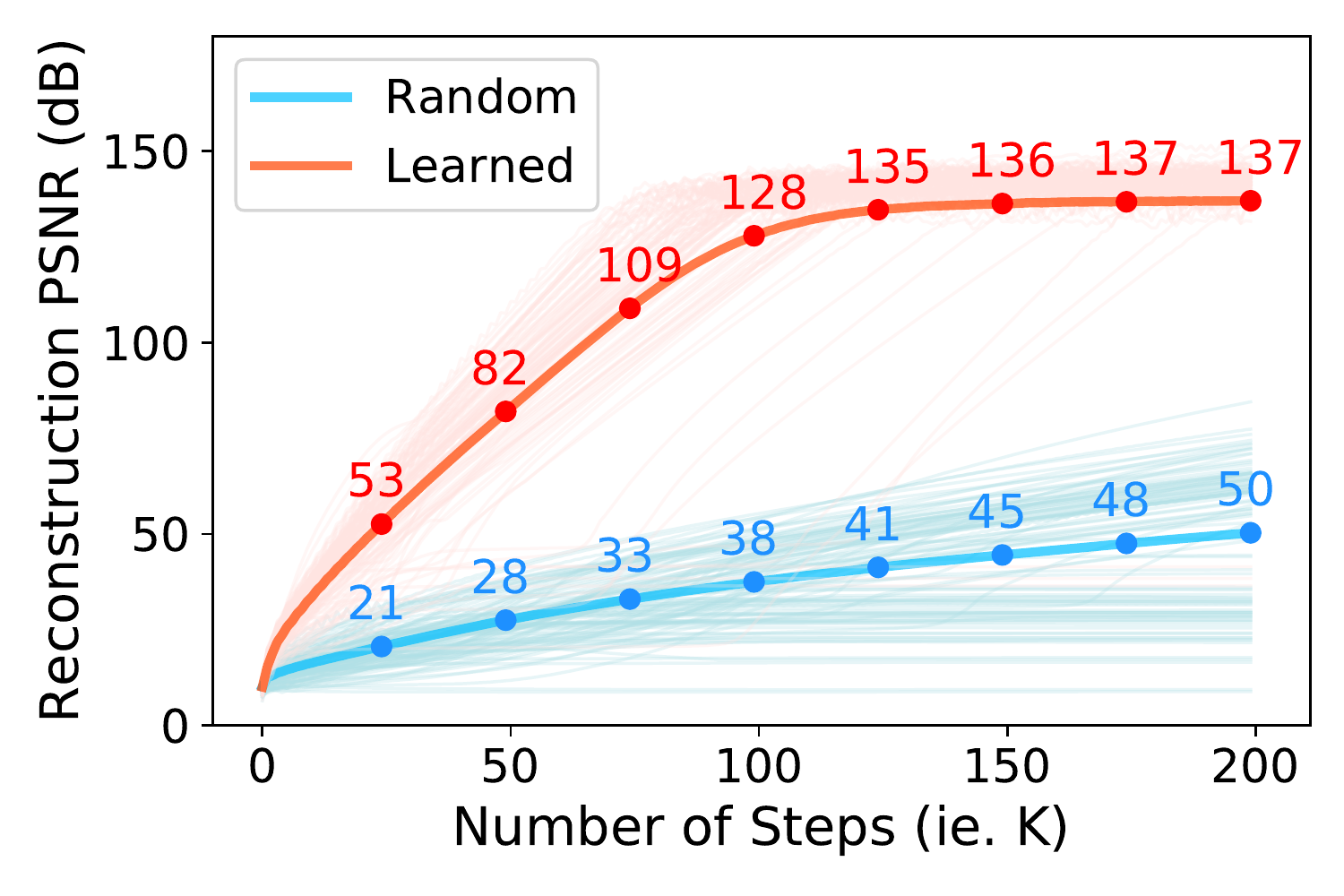}
    \caption{CIFAR10}
    \end{subfigure}
    \hfill
    \begin{subfigure}[t]{0.24\linewidth}
    \includegraphics[width=1\textwidth]{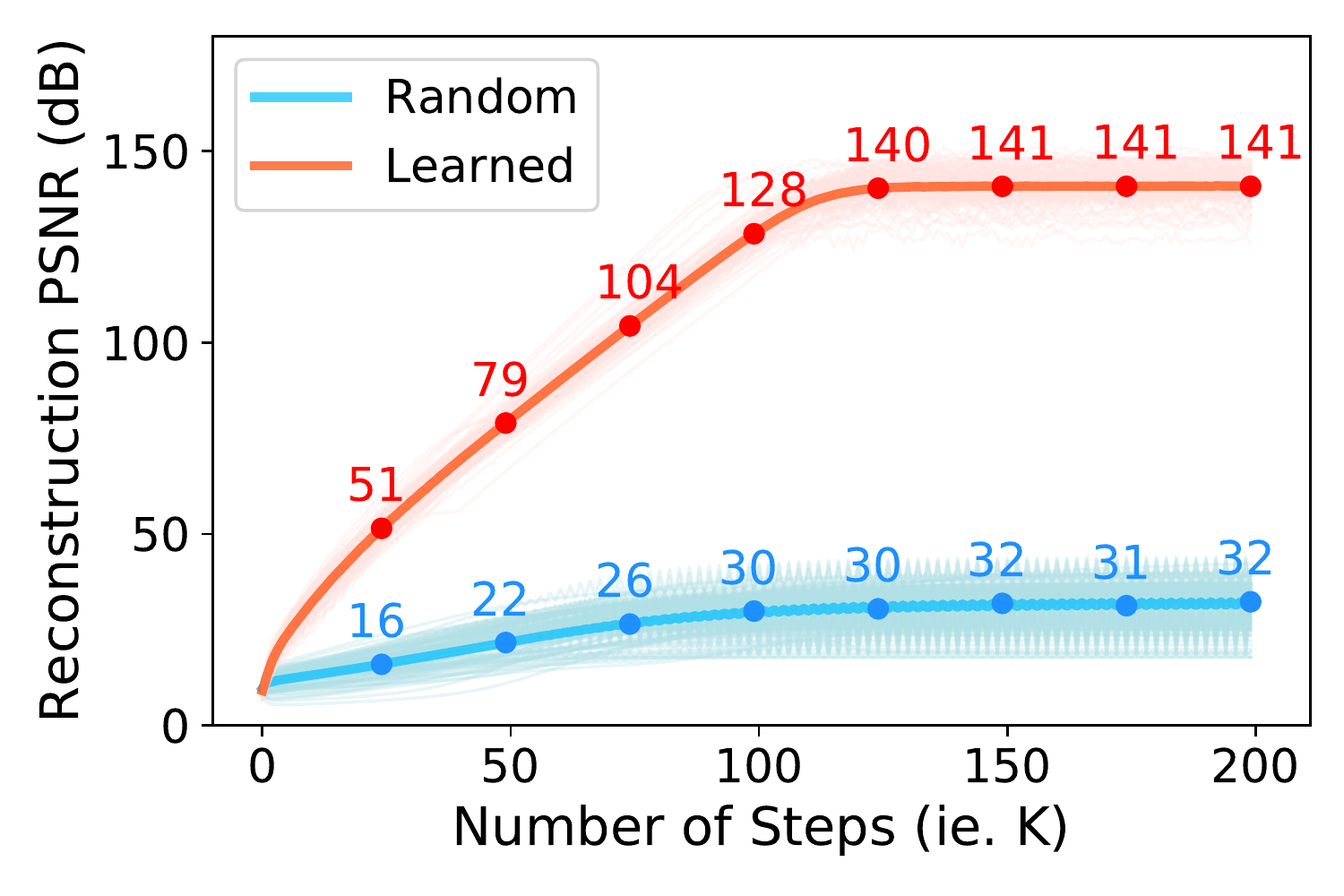}
    \caption{CelebA}
    \end{subfigure}
    \caption{Comparison of the reconstruction quality with  random and learned illumination patterns for different values of $K=1,\ldots,200$. We plot the average PSNR in bright color and the PSNR of randomly selected 100 samples in light shadows. \textbf{Learned} represents the reconstruction PSNR with learned illumination patterns (shown in red), and \textbf{Random} represents PSNR for random illumination patterns (shown in blue). The number of illumination patterns is $T=4$. Random illumination patterns are selected best out of 30 trials. The learned illumination patterns are trained on 128 training images and number of iterations $K=50$ during training.}
    \label{fig:diff_k}
\end{figure}

\begin{figure*}[!ht]
\centering 
\begin{subfigure}[t]{0.3\linewidth}
\includegraphics[width=1\linewidth]{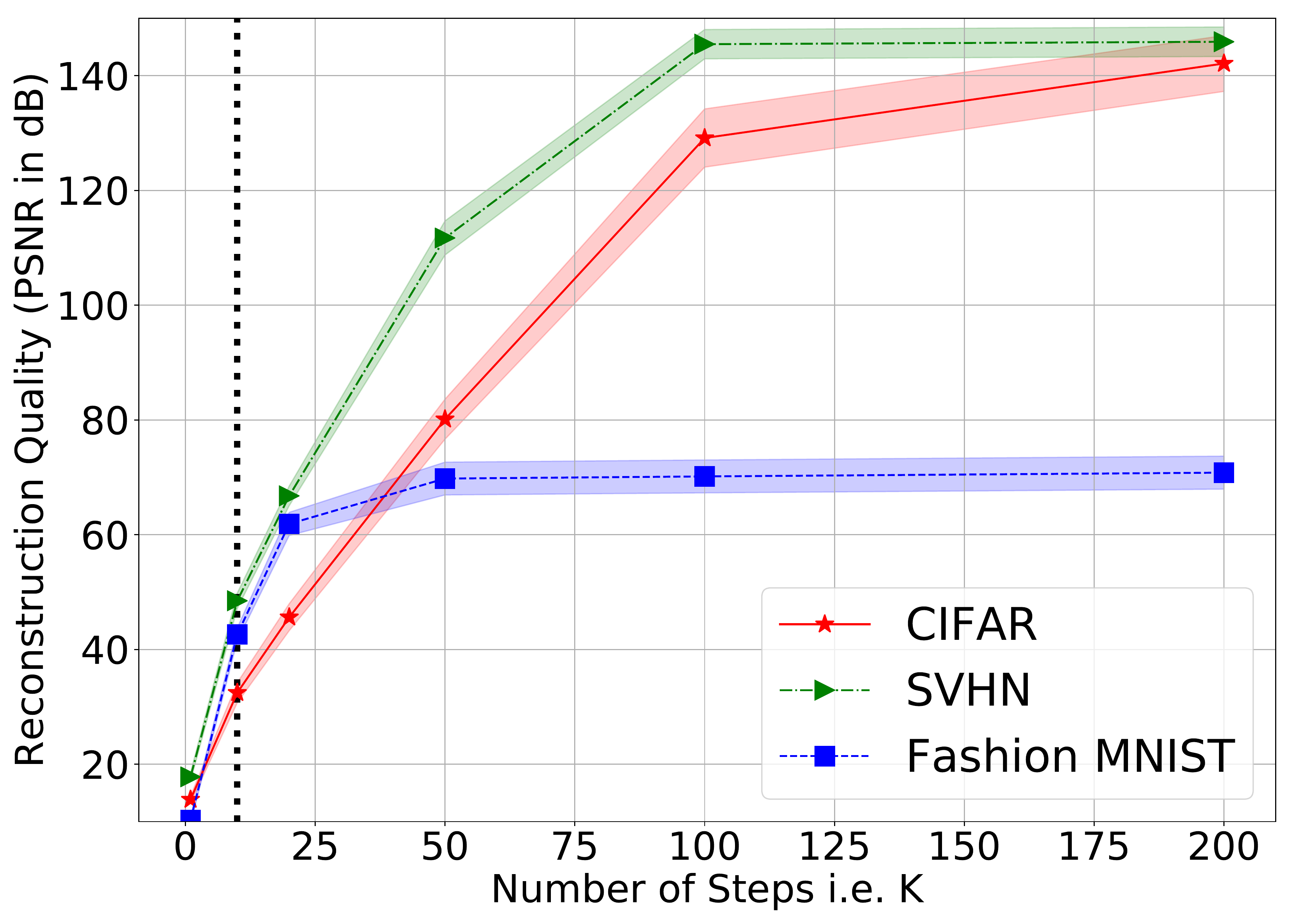}
\caption{Training K=10}
\end{subfigure}
\hfill
\begin{subfigure}[t]{0.3\linewidth}
\includegraphics[width=1\linewidth]{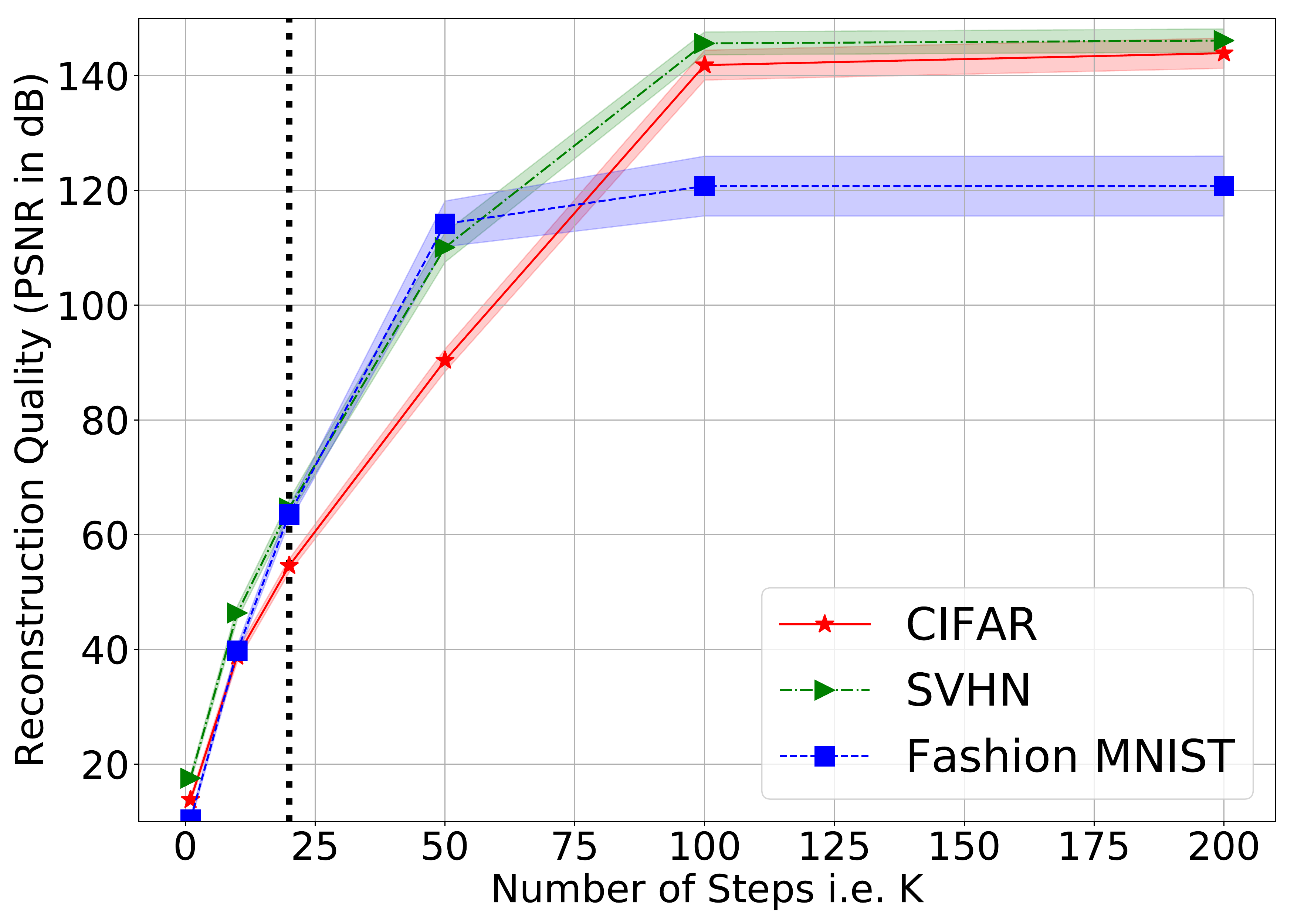}
\caption{Training K=20}
\end{subfigure}
\hfill
\begin{subfigure}[t]{0.3\linewidth}
\includegraphics[width=1\textwidth]{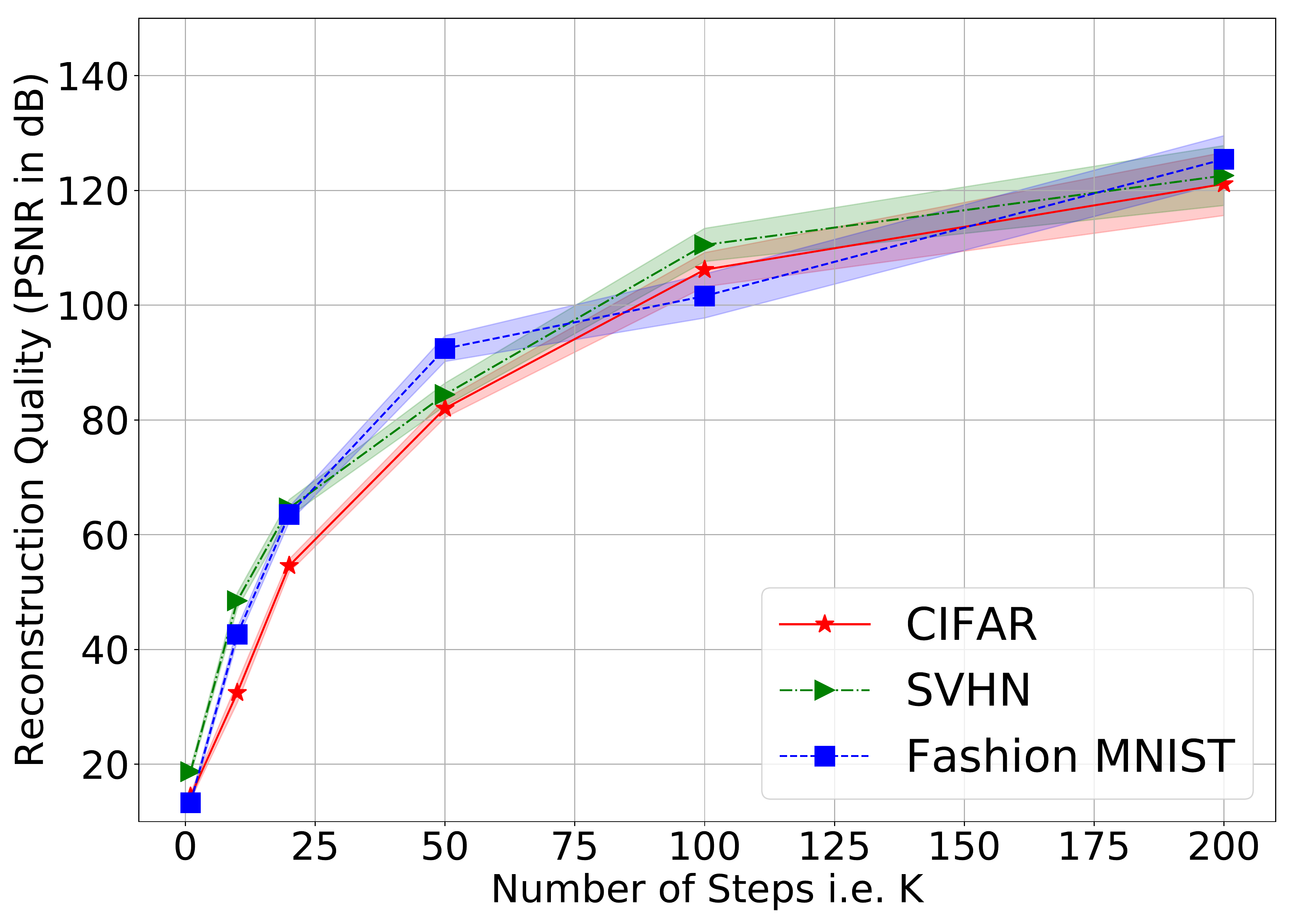}
\caption{Training K=Test K}
\end{subfigure}
\caption{Reconstruction quality vs number of iterations (layers) at test time (i.e., $K$ is different for training and testing with $T=4$). We show error bar of $\pm 0.25\sigma$ for each dataset. In (a) and (b), we fixed K (K=10, 20) and tested using different K. In (c), we trained and tested using the same number of layers.}
\label{fig:diff_k_train_test}
\end{figure*}

Figure~\ref{fig:diff_k} shows the performance of the learned and random illumination patterns as we increase $K$ to 200 at test time using the patterns learned for $K=50$. We observed that, with the learned patterns, the image reconstruction process converges faster and is more stable (smaller variance) than the case with random patterns. The  red curve in Fig.~\ref{fig:diff_k} has a steeper slope and narrower shades. Besides the default setting for $K=50$, we also learn the illumination patterns for different values of $K$. 
Figure~\ref{fig:diff_k_train_test} shows that we can recover images in a small number of iterations if we use learned illumination patterns. 
We also observe that we can perform better if we use more iterations in testing than in training. We have chosen $K=50$ for most of the experiments as a trade-off between computational cost and reconstruction performance.

\subsection{Generalization of learned patterns on different datasets}

To explore the generalizability of our learned illumination patterns, we use patterns learned on one dataset to recover images from another. The results are shown in Table.~\ref{table:generalzation}. As we can see in the table, the diagonal numbers are generally the best, and off-diagonal numbers are generally better than the ones with random illumination patterns. 

\begin{figure}[!t]
    \centering
    \includegraphics[width=0.6\textwidth]{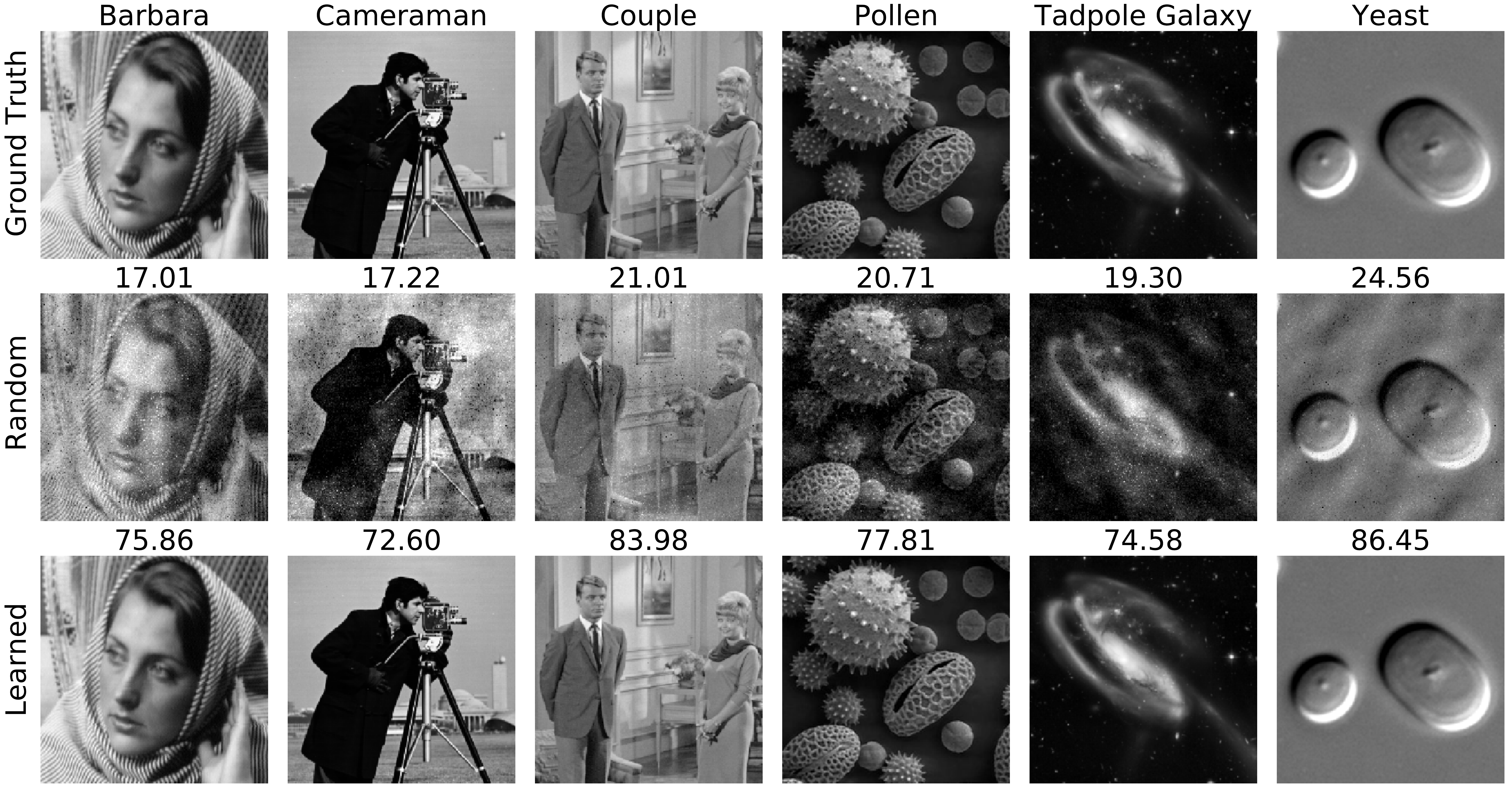}
    \caption{\textbf{First Row:} Ground truth images from image processing standard test datasets. \textbf{Second Row:} Reconstruction using random illumination patterns with uniform random distribution [0, 1] (we selected $T=4$ patterns that provided best results on celebA test images in \textbf{30 trials}).  PSNR numbers are shown on the top of reconstructed  images. \textbf{Third Row:} Reconstruction using the patterns trained on celebA dataset. Each image has $200\times 200$ pixels and the number of illumination patterns is $T=4$.}
    \label{fig:classical}
\end{figure}

We have also tested the learned illumination patterns on several classical images. Some results are shown in Fig.~\ref{fig:classical}. We used illumination patterns learned on 128 celebA images, but we can see that the learned illumination patterns perform better than the randomly chosen illumination patterns for classical images which indicates some generalizability of our learned illumination patterns.

\begin{table*}[]
\vspace{2mm}
\caption{Reconstruction performance of illumination patterns learned and tested on different datasets. Every column corresponds to patterns learned on a fixed dataset and tested on all. Rand column reports the performance of random illumination patterns.}
\vspace{-3mm}
\label{table:generalzation}
\begin{center}
\begin{small}
\begin{tabular}{@{}ccccccccccc@{}}
\toprule
 &
  \multicolumn{5}{c}{4 Illumination Patterns} &
  \multicolumn{5}{c}{8 Illumination Patterns} \\ \cmidrule(l){2-11} 
\multirow{-2}{*}{\begin{tabular}[c]{@{}c@{}}Test\\ \textbackslash{}Train\end{tabular}} &
  MNIST &
  FMNIST &
  CIFAR &
  SVHN &
  {\color[HTML]{656565} Rand} &
  MNIST &
  FMNIST &
  CIFAR &
  SVHN &
  {\color[HTML]{656565} Rand} \\ \midrule
MNIST &
  {\color[HTML]{000000} 101.50} &
  {\color[HTML]{000000} 83.13} &
  {\color[HTML]{000000} 78.04} &
  {\color[HTML]{000000} 54.83} &
  {\color[HTML]{656565} 28.61} &
  {\color[HTML]{000000} 111.79} &
  {\color[HTML]{000000} 92.66} &
  {\color[HTML]{000000} 82.75} &
  {\color[HTML]{000000} 74.86} &
  {\color[HTML]{656565} 46.43} \\
FMNIST &
  {\color[HTML]{000000} 64.47} &
  {\color[HTML]{000000} 92.44} &
  {\color[HTML]{000000} 85.14} &
  {\color[HTML]{000000} 76.41} &
  {\color[HTML]{656565} 30.05} &
  {\color[HTML]{000000} 82.37} &
  {\color[HTML]{000000} 108.85} &
  {\color[HTML]{000000} 85.80} &
  {\color[HTML]{000000} 93.09} &
  {\color[HTML]{656565} 52.05} \\
CIFAR &
  {\color[HTML]{000000} 31.77} &
  {\color[HTML]{000000} 46.79} &
  {\color[HTML]{000000} 82.02} &
  {\color[HTML]{000000} 68.12} &
  {\color[HTML]{656565} 27.51} &
  {\color[HTML]{000000} 52.75} &
  {\color[HTML]{000000} 84.94} &
  {\color[HTML]{000000} 104.61} &
  {\color[HTML]{000000} 87.85} &
  {\color[HTML]{656565} 60.64} \\
SVHN &
  {\color[HTML]{000000} 40.04} &
  {\color[HTML]{000000} 55.31} &
  {\color[HTML]{000000} 87.77} &
  {\color[HTML]{000000} 84.43} &
  {\color[HTML]{656565} 26.24} &
  {\color[HTML]{000000} 65.88} &
  {\color[HTML]{000000} 100.00} &
  {\color[HTML]{000000} 104.69} &
  {\color[HTML]{000000} 109.25} &
  {\color[HTML]{656565} 65.33} \\ \bottomrule
\end{tabular}

\end{small}
\end{center}

\end{table*}

\subsection{Robustness to noise}
\begin{figure}[!t]
\centering 
\begin{subfigure}[t]{0.35\linewidth}
\includegraphics[width=1\textwidth]{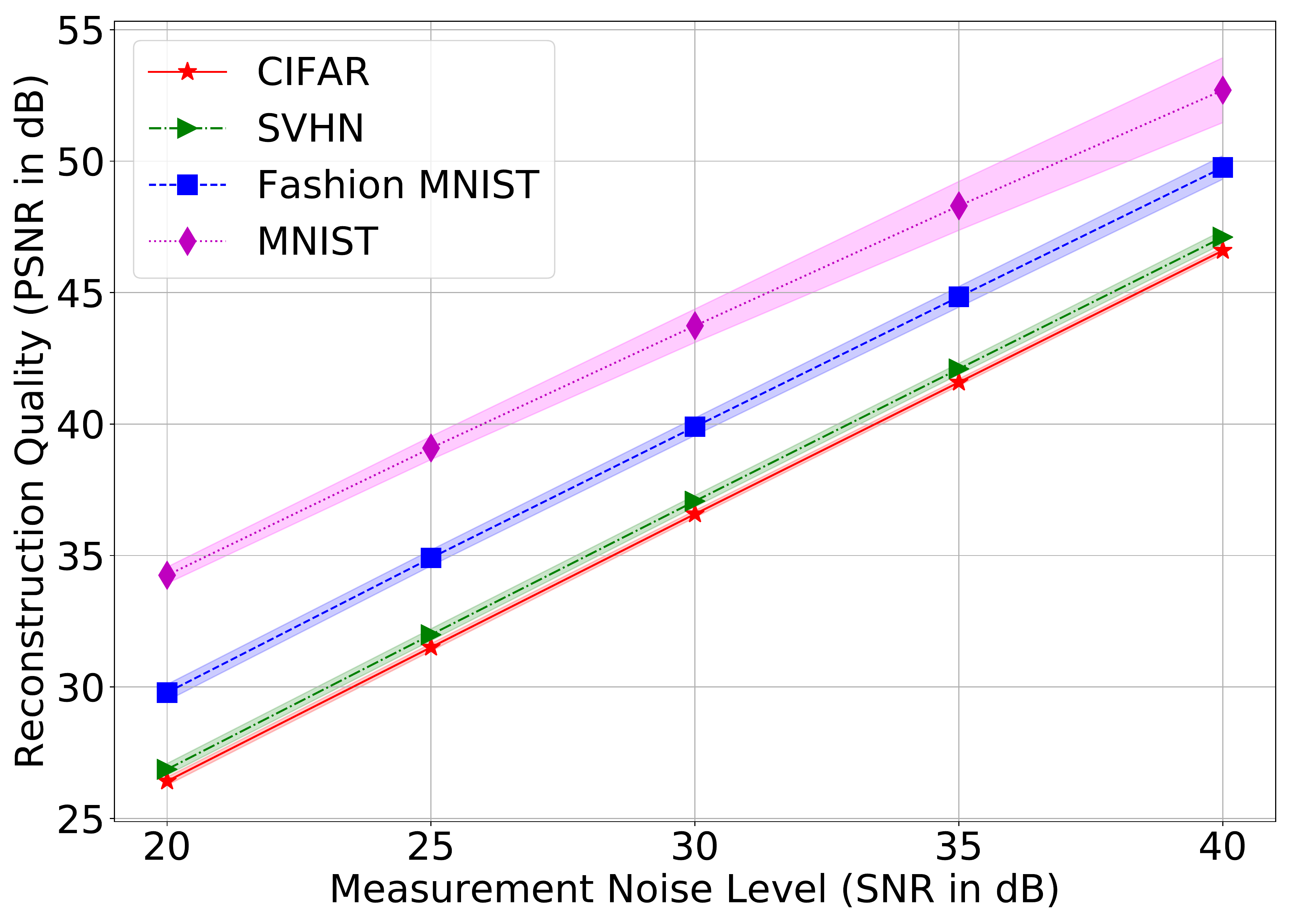}
\caption{Gaussian}
\end{subfigure}
\centering 
\begin{subfigure}[t]{0.35\linewidth}
\includegraphics[width=1\textwidth]{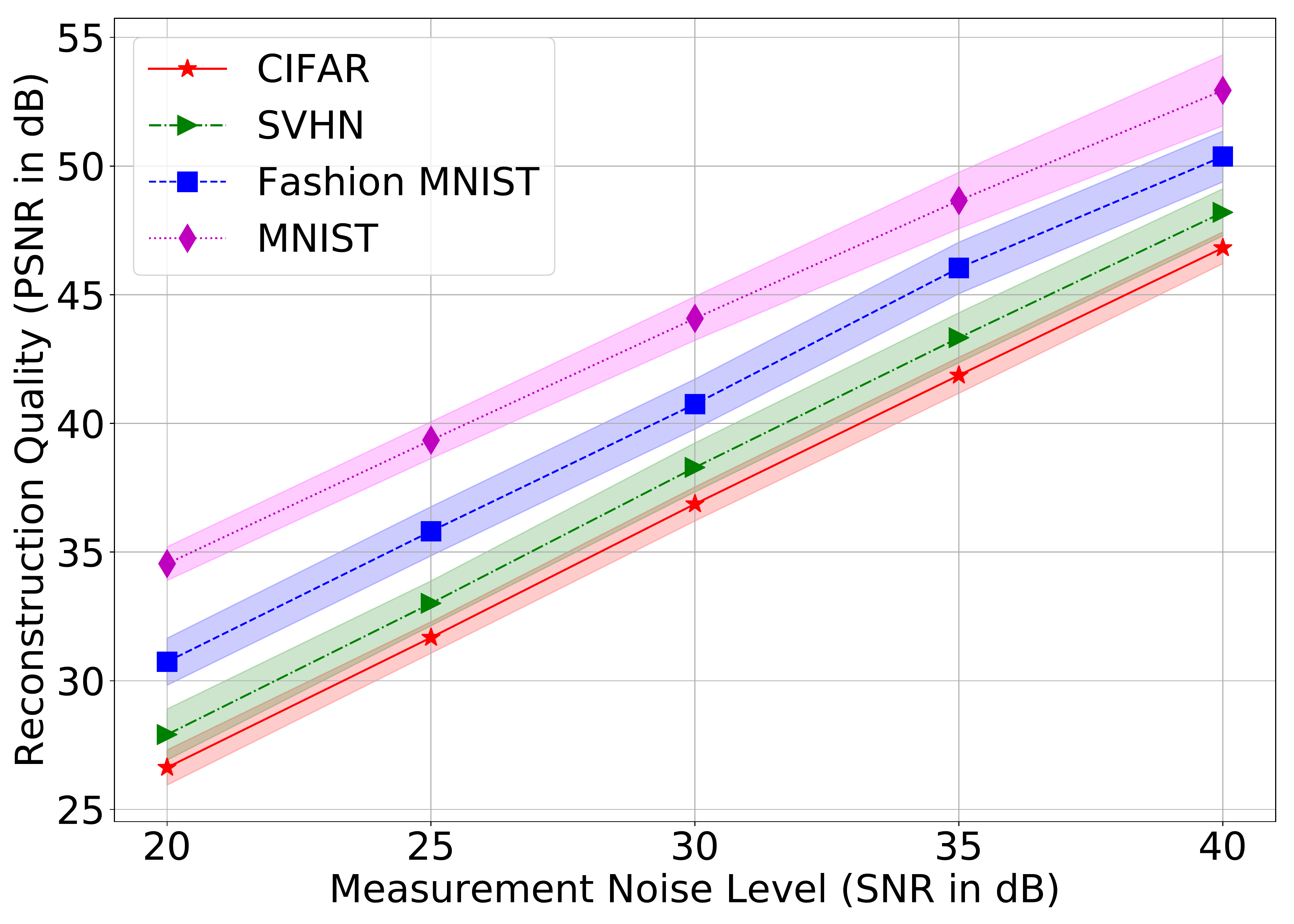}
\caption{Poisson}
\end{subfigure}
\caption{Reconstruction quality of the test images vs noise level of the measurements for different datasets. Here we show shaded error bar of  $\pm 0.25\sigma$ for each dataset. We learn the illumination patterns ($T=4$) on 128 noiseless training images of corresponding datasets.}
\label{fig:test_noise}
\end{figure}

To investigate the robustness of our method to noise, we train our illumination patterns on noiseless measurements obtained from the training datasets. We then added Gaussian and Poisson noise at different levels to the measurements from the test datasets. Poisson noise or shot noise is the most common in the imaging systems. We model the Poisson noise following the approach in \cite{metzler2018prdeep}. Let us denote the $i^{th}$ element of measurement vector corresponding to $t^{th}$ illumination pattern, $y_t$ as
\begin{equation}
    y_t(i)=|z_t(i)|+ \eta_t (i) \;\;\;\forall i=1,2,\ldots, m
\end{equation}
where $ \eta_t (i) \sim \mathcal{N}(0,\lambda|z_t(i)|)$ and $z_t=\mathcal{F}(d_t\odot x)$. We varied $\lambda$ to generate noise at different signal-to-noise ratios. Poisson noise affects larger values in measurements with higher strength than the smaller values. Since the sensors can measure only positive measurements, we kept the measurements positive by applying ReLU function after noise addition. We expect the reconstruction to be affected by noise as we did not use any denoiser. We observe the effect of noise in Figure~\ref{fig:test_noise}. Even though noise affects the reconstructions, we can get reasonable reconstruction up to a certain level of noise. The relationship between noise level and reconstruction performance also indicates that our phase retrieval system is quite stable.

\begin{table}[]
\caption{Reconstruction PSNR and run time (sec) of our and existing methods on different datasets.}
\label{table:compare_existing_methods}
\vskip 0.15in
\begin{center}
\begin{small}
\begin{tabular}{ccccccccccc}
\hline
\multirow{2}{*}{Algorithm} &
  \multicolumn{2}{c}{MNIST} &
  \multicolumn{2}{c}{F. MNIST} &
  \multicolumn{2}{c}{CIFAR10} &
  \multicolumn{2}{c}{SVHN} &
  \multicolumn{2}{c}{CelebA} \\ \cline{2-11} 
            & PSNR  & Time & PSNR  & Time & PSNR  & Time & PSNR  & Time & PSNR  & Time  \\ \hline
Fienup      & 13.17 & 0.31 & 23.28 & 0.30 & 32.79 & 0.29 & 40.57 & 0.29 & 31.57 & 5.42  \\
GS          & 12.96 & 0.31 & 23.28 & 0.30 & 32.71 & 0.29 & 40.99 & 0.29 & 31.69 & 5.22  \\
WirtFlow    & 10.74 & 0.07 & 13.72 & 0.06 & 16.28 & 0.05 & 19.65 & 0.04 & 16.93 & 1.26  \\
AmpFlow     & 17.09 & 0.07 & 32.83 & 0.05 & 42.47 & 0.05 & 47.43 & 0.04 & 37.76 & 1.24  \\
Kaczmarz    & 11.50 & 0.05 & 13.46 & 0.04 & 15.31 & 0.03 & 18.39 & 0.03 & 16.63 & 0.94  \\
Deep Models & 31.73 & 8.41 & 22.33 & 8.45 & 25.63 & 8.40 & 27.81 & 8.25 & 22.31 & 10.55 \\
Ours &
  \textbf{101.50} &
  \textbf{0.02} &
  \textbf{92.44} &
  \textbf{0.02} &
  \textbf{82.02} &
  \textbf{0.02} &
  \textbf{84.43} &
  \textbf{0.02} &
  \textbf{78.98} &
  \textbf{0.04} \\ \hline
\end{tabular}
\end{small}
\end{center}
\end{table}

\subsection{Comparision with existing methods}
We compare our method with various existing methods on different datasets and show the reconstruction PSNR (dB) and run time (seconds) per image in Table.~\ref{table:compare_existing_methods}. The description of the datasets can be found in section \ref{section:dataset}. We compare with Fineup  \cite{fienup1978reconstruction}, Gerchberg-Saxton  \cite{gerchberg1972practical}, Wirtinger Flow  \cite{candes2015phase}, Amplitude Flow  \cite{chen2015solving}, Kaczmarz  \cite{wei2015solving} and deep generative models \cite{metzler2020deep}. For \cite{fienup1978reconstruction,gerchberg1972practical,candes2015phase,chen2015solving,wei2015solving}, we use the code collected in PhasePack \cite{chandra2017phasepack}. In PhasePack, we restricted all the illumination patterns in the range of $[0,1]$ and set the maximum number of iterations to be $50$ which is the same as our default setting. We note that all of the methods in PhasePack use optimal spectral initialization; in contrast, we only use zero initialization. For deep generative models, we use a modified version of the publicly available code for \cite{metzler2020deep}. The code only provided pretrained DCGAN models for MNIST and F. MNIST; therefore, we trained our DCGAN models on the other datasets. This method is noticeably time-consuming because it optimizes over the latent vector for the deep model and uses 2000 iterations for each image where each iteration requires a forward and backward pass through the deep model. 
For all of the methods, we selected the best PSNR from 5 trials and report the average time. For our method, we report the run time with batch size equal to 1, which aligns with other methods.
We observe from the table that our method with learned patterns performs significantly better than all the other algorithms in terms of both reconstruction quality and run time.

\bibliography{learnPR}
\bibliographystyle{IEEEtran}

\end{document}